# Transport evidence of asymmetric spin-orbit coupling in few-layer superconducting 1Td-MoTe$_2$

Jian Cui[1†], Peiling Li[1,2†], Jiadong Zhou[3†], Wen-Yu He[4†], Xiangwei Huang[1,2], Jian Yi[5], Jie Fan[1], Zhongqing Ji[1], Xiunian Jing[1,6], Fanming Qu[1], Zhi Gang Cheng[1], Changli Yang[1,6], Li Lu[1,6], Kazu Suenaga[7], Junwei Liu[4], Kam Tuen Law[4], Junhao Lin[8*], Zheng Liu[3,9,10*] and Guangtong Liu[1*]

[1]Beijing National Laboratory of Condensed Matter Physics, Institute of Physics, Chinese Academy of Sciences, Beijing 100190, China

[2]University of Chinese Academy of Sciences, Beijing 100049, China

[3]Centre for Programmable Materials, School of Materials Science and Engineering, Nanyang Technological University, Singapore 639798, Singapore

[4]Department of Physics, Hong Kong University of Science and Technology, Clear Water Bay, Hong Kong, China

[5]Ningbo Institute of Industrial Technology, Chinese Academy of Sciences, Ningbo 315201, China

[6]Collaborative Innovation Center of Quantum Matter, Beijing 100871, China

[7]National Institute of Advanced Industrial Science and Technology (AIST), Tsukuba 305-8565, Japan

[8]Department of Physics, Southern University of Science and Technology, Shenzhen 518055, China

[9]Centre for Micro-/Nano-electronics (NOVITAS), School of Electrical & Electronic Engineering, Nanyang Technological University, 50 Nanyang Avenue, Singapore 639798, Singapore

[10]CINTRA CNRS/NTU/THALES, UMI 3288, Research Techno Plaza, 50 Nanyang Drive, Border X Block, Level 6, Singapore 637553, Singapore

† These authors contributed equally to this work. Correspondence and requests for materials should be addressed to J.L (email:lin.junhao.stem@gmail.com), Z.L. (email: z.liu@ntu.edu.sg) and G.L. (email:gtliu@iphy.ac.cn)

**Abstract**

Two-dimensional (2D) transition metal dichalcogenides (TMDCs) $MX_2$ (M=W, Mo, Nb, and X=Te, Se, S) with strong spin-orbit coupling (SOC) possess plenty of novel physics including superconductivity. Due to the Ising SOC, monolayer $NbSe_2$ and gated $MoS_2$ of 2H structure can realize the Ising superconductivity phase, which manifests itself with in-plane upper critical field far exceeding Pauli paramagnetic limit. Surprisingly, we find that a few-layer 1Td structure $MoTe_2$ also exhibits an in-plane upper critical field ($H_{c2,\|}$) which goes beyond the Pauli paramagnetic limit. Importantly, the in-plane upper critical field shows an emergent two-fold symmetry which is different from the isotropic $H_{c2,\|}$ in 2H structure TMDCs. We show that this is a result of an asymmetric SOC in 1Td structure TMDCs. The asymmetric SOC is very strong and estimated to be on the order of tens of meV. Our work provides the first transport evidence of a new type of asymmetric SOC in TMDCs which may give rise to novel superconducting and spin transport properties. Moreover, our findings mostly depend on the symmetry of the crystal and apply to a whole class of 1Td TMDCs such as 1Td-$WTe_2$ which is under intense study due to its topological properties.

In conventional BCS singlet superconductor, the external magnetic field becomes detrimental to the superconductivity state through orbital depairing effect and Pauli paramagnetism[1]. In the 2D atomically thin superconductor, the orbital effect is suppressed and Pauli paramagnetism plays the dominant role when an in-plane magnetic field is applied. The effect of in-plane magnetic field on the 2D superconductor is recently studied in the superconducting 2H type transition metal dichalcogenides (TMDCs), including gated $MoS_2$[2,3], 2D $NbSe_2$[4,5], and monolayer $TaS_2$[6]. Interestingly, the in-plane upper critical field ($H_{c2,\|}$) is observed to be strongly enhanced beyond the Pauli paramagnetic limit[7,8]. The large enhancement of $H_{c2,\|}$ in the 2H TMDCs originates from the strong Ising SOC due to the breaking of an in-plane mirror symmetry and the presence of the out-of-plane mirror symmetry in the crystal structure. As electron spins are pinned to the out-of-plane directions, this phenomenon is named Ising superconductivity. Ising superconductivity[2-4] with its promising applications in equal spin Andreev reflections[9], proximity phenomenon[10], engineering Majorana fermions[9,11,12], and topological superconductivity[13,14] has sparked intense research interest in condensed matter physics. So far, the study of SOC effect on superconductivity in TMDCs is limited to the Ising SOC. In this work, we systematically study the superconducting few-layer 1Td-MoTe$_2$. A few-layer 1Td-MoTe$_2$, unlike its 2H structure counterparts, breaks both the in-plane mirror symmetry and out-of-plane mirror symmetry. We show that the resulting SOC is asymmetric in the three spatial directions. By combining the low-temperature transport measurements and self-consistent mean-field calculations, we demonstrate that the in-plane upper critical field in the superconducting few-layer 1Td-MoTe$_2$ exceeds the Pauli limit in the whole in-plane directions. Importantly, we theoretically predicted and experimentally verified that the in-plane upper critical field shows an emergent two-fold symmetry due to the new type of anisotropic SOC. From the experimental data, we further estimated that the SOC strength is on the orders of tens of meV which is also consistent with the results of our first-principle calculations. Our work gives clear evidence that anisotropic SOC plays an important role in determining the properties of superconductivity in MoTe$_2$.

In our experiment, the high crystalline few-layer MoTe$_2$ crystals were produced by molten-salt assisted chemical vapor deposition (CVD) method[15,16] (see details in the *Materials and Methods* section and Supplementary Figure S1). The optical images of the as-synthesized MoTe$_2$ layers with different thicknesses are shown in Fig. 1a. Similar to our previous results[15], the mono- and few-layer MoTe$_2$ can have a size up to 100 μm with a rectangular shape. Figure 1b shows the Raman spectra of the as-synthesized MoTe$_2$ with different layers, where we ascribe the Ag modes at 127, 161, and 267 cm$^{-1}$ to 1Td-MoTe$_2$, which is further supported by the following STEM measurements. Note that the Ag mode located at 267 cm$^{-1}$ shows a blue-shift with increasing sample thickness, which is similar to the Raman shift in other 2D materials such as MoS$_2$[17] and WS$_2$[18].

The atomic structure of few-layer MoTe$_2$ is further characterized by annular dark-field (ADF) scanning transmission electron microscopy imaging (STEM). Figure 1c shows the atom-resolved STEM image of few-layer MoTe$_2$ in large scale (see Supplementary Figure S2a, b for the chemical purity verified by energy dispersive X-ray spectra). The 1Td phase and 1T' phase share the same in-plane crystal structure (Supplementary Figure S2c), but the two structures have different stacking. The 1T' crystallizes in monoclinic shape and keeps the global inversion center, while the 1Td phase has the vertical stacking and belongs to the non-centrosymmetric space group Pmn2$_1$, as shown by the atomic models in Fig. 1c. Therefore, the projection of the scattering potential is different in these two phases and can be distinguished by their STEM images. At room temperature, bulk MoTe$_2$ usually crystalized in the monoclinic 1T' phase. By comparing the simulated images both in 1Td and 1T' phase shown in Fig. 1c, we unambiguously found that the few-layer MoTe$_2$ is in the 1Td phase rather than the bulk 1T' phase at room temperature. This is different from the previous reports[19,20] where the 1Td phase only occurs at temperature below 200 K. Additionally, the temperature dependence of Raman spectra and sheet resistance shown in Fig. S3 also confirm the few-layer MoTe$_2$ is in the Td-phase, i.e., no phase transition is observed as lowering down the temperature. Therefore, the 1Td phase could be the intrinsic feature of the CVD-grown few-layer MoTe$_2$, which is presumably caused by the reduced thickness of MoTe$_2$[21].

Figure 1d shows the temperature dependence of the normalized four-terminal sheet resistance ($R/R_{300\,K}$), measured at zero magnetic field, for MoTe$_2$ films with thickness from 2 nm to 30 nm (See Supplementary Figure S4 for raw data). At high temperatures, all samples measured show a metallic behavior with d$R$/d$T$ > 0, indicating that the phonon scattering dominates the transport. As the temperature is further lowered, the samples enter a disorder-limited transport regime prior to the eventual superconducting state. The residual resistance ratio, RRR = $R_{300K}/R_n$ with $R_{300K}$ the room-temperature sheet resistance and $R_n$ the normal-state sheet resistance right above the superconducting transition, varies from 1.15 of the 2.0-nm-thick to 2.33 of the 30-nm-thick MoTe$_2$ crystals (Supplementary Table 1). However, the largest sheet resistance $R_n = R_{5K} = 810\ \Omega\,\square^{-1}$ found in the 2.0-nm-thick device is much smaller than the quantum resistance $R_Q = h/(2e)^2 = 6{,}450\ \Omega$, suggesting that our samples are all in the low-disorder regime[22].

At low temperatures, superconductivity is observed for all samples. To examine the thickness-dependent superconductivity, in the inset of Fig. 1d we show the temperature dependence of the reduced resistance, $r = R/R_n = R/R_{5K}$, in a low-temperature regime ($T \leq 5.5$ K) for samples with different thickness. Empirically, critical transition temperatures for the superconductivity, $T_{c,r}$, can be extracted from the $R$ vs $T$ curve. This is realized by picking up the points firstly encountered with the predefined reduced resistance $r$ from the normal state into the superconducting state. Such transition temperatures, extracted at typical values $r$=0, 0.5，and 0.9, are listed in supplementary Table S1 for our samples with different thickness. It is found that, $T_{c,0}$ increase from 0.35 K to 3.16 K with increasing sample thickness from 2 nm to 30 nm, and the $T_{c,0}$ values of our samples are surprisingly higher than $T_{c,0} \sim 0.1$ K as reported in stoichiometric bulk MoTe$_2$[20]. In bulk MoTe$_2$, Te-vacancy-enhanced superconductivity has been previously reported[16] with the highest $T_{c,0} \sim 1.3$ K, which is still much lower than $T_{c,0} = 3.16$ K observed in our 30-nm-thick MoTe$_2$ crystals. In addition, a significant broadening on superconducting transition are observed for 2-nm-thick device, which can be attributed to the enhanced thermal fluctuations in two dimensions[1,23]; similar behaviors have been observed in few-layer Mo$_2$C[24] and NbSe$_2$[4,25] superconductors reported recently.

The few-layer MoTe$_2$ crystals provide an ideal platform to study their transport properties in the 2D limit. To investigate the dimensionality of the superconductivity in few-layer MoTe$_2$, we firstly studied the temperature dependence of the upper critical magnetic field $\mu_0 H_{c2}$, which is defined as the magnetic field corresponding to a predefined reduced resistance $r=R/R_n=0.5$. Figure 2a,b show the superconducting resistive transitions of a 8.6-nm-thick MoTe$_2$ device with the magnetic field perpendicular and parallel to the sample surface, respectively, measured at fixed temperatures. In both cases, one can see that the superconducting transition shifts gradually to lower magnetic fields with the increase of temperature. The temperature-dependent upper critical fields in directions parallel and perpendicular to the sample surface, denoted by $\mu_0 H_{c2,\parallel}$ and $\mu_0 H_{c2,\perp}$ respectively, are plotted in Fig. 2c. We found that the superconductivity was more susceptible to perpendicular magnetic fields than to parallel magnetic fields, and a large ratio of $H_{c2,\parallel}/H_{c2,\perp} \approx 7$ is obtained in the 8.6-nm-thick sample, indicating a strong magnetic anisotropy. This is true for all samples, and the ratio reaches up to 26 for 2.7-nm-thick sample (see Supplementary Figure S5 for more samples). A linear temperature dependence was observed for $H_{c2,\perp}$, which can be well fitted by the phenomenological 2D Ginzburg-Landau (GL) theory[1],

$$H_{c2,\perp}(T) = \frac{\phi_0}{2\pi\xi_{GL}^2}\left(1 - \frac{T}{T_{c,0}}\right) \qquad (1)$$

where $\xi_{GL}$ is the zero-temperature GL in-plane coherence length and $\phi_0$ is the magnetic flux quantum, as shown by the blue dashed line in Fig. 2c that gives $\xi_{GL} = 20.79$ nm (See Supplementary Table S1 for more data on samples with different thickness). Compared with the sample thickness of 8.6 nm as measured by AMF (Supplementary Figure S5f), the coherence length $\xi_{GL} = 20.79$ nm is approximately two times of the thickness, indicating that the depairing effect from the orbital magnetic field is strongly suppressed. As a result, the in-plane magnetic field induced paramagnetism determines $H_{c2,\parallel}(T)$.

The 2D behavior of the superconducting few-layer MoTe$_2$ is further confirmed by the experiments with tilted magnetic field. Figure 2d shows the magnetic field dependence of the sheet resistance $R$ under different $\theta$ at 0.3 K, where $\theta$ is the tilted angle between the normal of the sample plane and the direction of the applied magnetic field (the inset of Fig. 2c). Clearly, the superconducting transition shifts to higher field with the external magnetic field rotating from perpendicular $\theta=0º$ to parallel $\theta=90º$ (See Supplementary Figure S6 for more data on different MoTe$_2$ samples). The upper critical field $\mu_0 H_{c2}$ was extracted from Fig. 2d and plotted in Fig. 2e as a function of the tilted angle $\theta$. In Fig. 2e, a cusp-like peak is clearly observed at $\theta=90º$ where the external magnetic field is aligned in parallel to the sample surface, which is apparently sharper for thinner sample (Supplementary Figure S6). Curves fitted with the 2D Tinkham model[1] and 3D anisotropic Ginzburg-Landau model shows the data consistence with both models for $\theta<85º$ and $\theta>95º$, whereas for $85º <\theta<95º$ the cusp-shaped dependence can only be explained with the 2D Tinkham model. It shows that our superconducting few-layer MoTe$_2$ manifests the 2D nature of the superconductivity. For a 2D superconductor with $d_{sc} \ll \xi_{GL}$, the *V-I* dependence as a function of temperatures is measured and shown in Fig. 2f. The Berezinskii-Kosterlitz-Thouless temperature is estimated to be $T_{BKT} =1.47$K, only slightly larger than $T_{c,0}=$

1.4 K of the sample. With the above evidences, the 2D superconductivity is convincingly confirmed in our few-layer 1Td-MoTe$_2$ samples.

Now we turn to discuss the most important findings of our experiments, i.e., the observation of the in-plane upper critical field ($H_{c2,\|}$) beyond the Pauli limit and the emergent two-fold symmetry of $H_{c2,\|}$. In conventional Bardeen-Cooper Schrieffer (BCS) superconductors, sufficiently high external magnetic field can destroy the superconductivity by breaking Cooper pairs via the coexisting orbital[1,26] and Zeeman spin splitting effect[7,8]. For the few-layer sample, the orbital effect of the in-plane magnetic field is greatly suppressed due to the reduced dimensionality[1], and consequently $H_{c2,\|}$ is solely determined by the interaction between the external magnetic field and the spin of the electrons. When the magnetization energy gained from the applied magnetic field approaches the superconducting condensation energy, the Cooper pairs are broken and superconductivity is destroyed at the characteristic field given by the Clogston-Chandrasekhar[7] or Pauli paramagnetic limit $H_p = \sqrt{2}\Delta_0/(g\,\mu_B)$, where $\Delta_0$=1.76$k_B T_c$, $g$ as the $g$ factor and $\mu_B$ as the Bohr magneton. The observation of the $H_{c2,\|}$ in our few-layer MoTe$_2$ is summarized in Fig. 3 for different samples. Figure 3a displays the superconducting transition in 1Td-MoTe$_2$ devices with various thicknesses under in-plane magnetic field measured at 0.3 K. Clearly, the superconductivity in 1Td-MoTe$_2$ can persist to higher in-plane magnetic field as the thickness is lowered down. From Figs. 3b and c, it can be seen that the values of $H_{c2,\|}$ for the six typical samples with different thicknesses are all larger than $H_p$, in marked contrast to their bulk counterpart that is well below the $H_p$[20]. More generally, the magnetic field dependence of the sheet resistance of a 3-nm thick MoTe$_2$ sample is further measured at $T$=0.3K ($T$=0.07$T_c$) with different in-plane tilted angle $\varphi$ as shown in Fig. 4a (see Supplementary Figure S8 for $T$ = 0.3 $T_c$, 0.6 $T_c$, and 0.95 $T_c$). Surprisingly, an emergent two-fold symmetry of $H_{c2,\parallel}$ has been observed in few-layer 1Td-MoTe$_2$ with the $H_{c2,\parallel}$ beyond $H_p$ in all the in-plane directions as shown in Fig. 4b. As the magnetic field tilted from y-axis ($\varphi = 0°$) to x-axis ($\varphi = 90°$) (the relation between the x- and y- axis and the crystal axis is shown in the Supplementary Fig. S9), we can see that the superconducting transition moves from higher field to lower field. From the low temperature ($T=0.07T_c$) to the temperature near $T_c$, the observed 2-fold symmetry $H_{c2,\parallel}$ stays robust. The emergent 2-fold symmetry observation in $H_{c2,\parallel}$ which exceeds $H_p$ at low temperatures is in sharp contrast with the standard BCS prediction and becomes highly nontrivial.

Similar anomalous enhancement of $H_{c2,\parallel}$ has been observed in layered superconductors in the dirty limit with strong spin-orbital coupling (SOC), which can be explained by spin–orbit scattering[27-32] (SOS) effect using the microscopic Klemm-Luther-Beasley (LKB) theory[31]. However, the SOS mechanism is inadequate to interpret our data because the samples are in the low-disorder regime as discussed earlier. It is known that inhomogeneous superconducting states, such as Fulde–Ferrell–Larkin–Ovchinnikov (FFLO) state[33-36] or helical state[37], can also enhance $H_{c2,\parallel}$, which have been observed in heavy Fermion superconductors[32], organic superconductors[38,39], and monolayer Pb films[40]. However, for superconductors induced by FFLO state, the theoretical value[38] of $H_{c2,\parallel}/H_p$ is in the range of 1.5~2.5, smaller than the observed value of 2.8 in our 2.7-nm-thick sample. Moreover, the FFLO characteristic upturn[33-35] of $H_{c2,\parallel}(T_c)$ at low temperature is missing in our experimental data as shown in Fig. 3c. Therefore, the FFLO state can be ruled out.

Recently in monolayer NbSe$_2$ and gated MoS$_2$ superconductor, the anomalous enhancement of $H_{c2,\parallel}$ beyond $H_p$ has been interpreted by the Ising SOC protected Ising superconductivity mechanism[2-4]. A monolayer TMDCs with 2H structure possesses an out-of-plane mirror symmetry, whereas the in-plane inversion symmetry is broken. The mirror symmetry restricts the crystal field ($\varepsilon$) to the plane, while the inversion symmetry breaking can induce strong SOC splitting, giving rise to an effective Zeeman-like magnetic field $H_{so}(k) \propto k \times \varepsilon$ (~100 T for gated $MoS_2$ and ~660 T for $NbSe_2$) with opposite out-of-plane direction at the K and –K valleys of the Brillouin zone[42]. Thus the electron spins are pinned along the out-of-plane directions, and they are antiparallel to each other for electrons with opposite momenta; that is to say, spins of electrons of Cooper pairs are polarized by the large out-of-plane effective Zeeman field and thus becomes insensitive to the external in-plane magnetic field, which results in the enhancement of in-plane $H_{c2,\parallel}$. However, in few-layer $MoTe_2$, the absence of out-of-plane mirror symmetry gives rise to a more complicated SOC field beyond the Ising SOC.

In order to fully understand the experimental results in the few-layer 1Td-$MoTe_2$, we focus on the 1Td-$MoTe_2$ bilayer and construct an effective model from the symmetry point of view. The crystal structure of bilayer 1Td-$MoTe_2$ is shown in Fig. S9 and it has the same symmetry properties as bulk crystals where only the mirror symmetry in the *y* direction is preserved, while both the out-of-plane mirror symmetry and the in-plane mirror symmetry in the *x* direction are broken. Since the bilayer 1Td-$MoTe_2$ respects the time reversal symmetry and the mirror symmetry in the *y* direction (Supplementary Figure S9), the spin-orbit coupling at the Fermi level is restricted to an effective form $H_{soc} = \boldsymbol{g} \cdot \boldsymbol{\sigma}$ with $\boldsymbol{g} = (x_1 \sin\varphi, y_1 \cos\varphi, z_1 \sin\varphi)$ in the first order (the complete form is derived in the Supplementary Information), where $\varphi$ is the polar angle for the Fermi wave vector. In the bilayer 1Td-$MoTe_2$, the breaking of in-plane mirror symmetry in the *x* direction generates the out-of-plane Ising component $g_z$, and the breaking of the out-of-plane mirror symmetry gives rise to the anisotropic in-plane components $(g_x, g_y)$ of the SOC. As a result, all the three components of SOC exist in the bilayer 1Td-MoTe$_2$. Similar to 2H structure TMDCs, $g_z$ strongly enhances in-plane $H_{c2}$. Interestingly, the in-plane anisotropic SOC will give rise to a two-fold symmetry in in-plane $H_{c2}$ as discussed below.

To quantitatively validate the asymmetric SOC enhanced upper critical field, we calculate the in-plane spin susceptibility (see Supplementary Figure S10 and the calculation procedure). At the zero temperature limit, the superconductivity normal metal transition driven by in-plane magnetic field occurs when $\frac{1}{2}N_0\Delta_0^2 + \frac{1}{2}\chi_s H_\parallel^2 = \frac{1}{2}\chi_N H_\parallel^2$, where the two sides of the equation correspond to the energy for the superconducting state and the normal state respectively. Here $\chi_s$ and $\chi_N$ denote the spin susceptibility of the superconducting state and the normal state respectively, and $N_0$ is the density of states at the Fermi level. In the presence of the spin-orbit coupling, the superconducting spin susceptibility has a finite value as is seen from Fig. 4c. The enhancement of the in-plane $H_{c2}$ by the spin-orbit coupling field becomes understandable since $H_{c2}(\varphi) = \sqrt{\frac{N_0}{\chi_N - \chi_s}}\Delta_0$. The two-fold angle dependence of $H_{c2}$ is also consistently explained by the anisotropic in-plane spin susceptibility $\chi_s$ shown in the inset of Fig. 4c. In Fig. 4b, the in-plane $H_{c2}(\varphi)$ at zero temperature with two-fold symmetry is plotted with the SOC field at the Fermi level $\boldsymbol{g} = k_B T_c (49\sin\varphi, 68\cos\varphi, 67\sin\varphi)$ and fits well with the experimental data measured at the temperature $T = 0.07\ T_c$. As $k_B T_c$ is about 0.345meV, the SOC field strengths are in the order of tens of meV which are in good agreement with the spin bands splitting shown from the first principle calculation (see Supplementary Figure S10). The mean-field calculations for the pairing order parameter dependence on the in-plane magnetic field along $x$ and $y$ directions are further carried out to obtain the magnetic field – temperature phase diagram as shown in Fig. 4d and e. The $H_{c2}$ along $x$ and $y$ directions has strong anisotropy from zero temperature to near $T_c$ and shows the same trend as the experimental data measured in high temperature in Fig. 4b.

In summary, for the first time, we demonstrate that the properties of the superconducting state of 1Td-MoTe$_2$ are strongly affected by a new type of asymmetric SOC which is in the order of tens of meV. Such strong SOC will create strong triplet pairing correlations in the material and may affect the pairing symmetry as well. Due to its large magnitude, the SOC may also have effects on the normal state spin transport of the system[43-45]. Importantly, the finding of the asymmetric SOC mostly depends on the symmetry of the crystal and similar asymmetric SOC are expected to exist in other 1Td structure TMDCs such as the recently well studied 1Td-WTe$_2$. Our findings on the new type of asymmetric SOC in 1Td-MoTe$_2$ are expected to promote further studies on the exotic superconducting and normal state phenomena in TMDCs.

**Materials and Methods**

**CVD synthesis of highly crystalline few-layer MoTe$_2$**. The few-layer MoTe$_2$ samples were synthesized via CVD method inside a furnace with a 1 inch diameter quartz tube. Specifically, one alumina boat containing precursor powder (NaCl: MoO$_3$=1:5) was put in the center of the tube. Si substrate with a 285 nm thick SiO$_2$ on top was placed on the alumina boat with polished side faced down. Another alumina boat containing Te powder was put on the upstream side of quartz tube at a temperature of about 450 $^o$C. Mixed gas of H$_2$/Ar with a flow rate of 15/80 sccm was used as the carrier gas. The furnace was ramped to 700 $^o$C at a rate of 50 $^o$C/min and held there for about 4 min to allow the growth of few-layer MoTe$_2$ crystals. After the reaction, the temperature was naturally cooled down to room temperature. All reagents were purchased from Alfa Aesar with purity exceeding 99%.

**Raman Characterization**. Raman measurements with an excitation laser of 532 nm were performed using a WITEC alpha 300R Confocal Raman system. Before the characterization, the system was calibrated with the Raman peak of Si at 520 cm$^{-1}$. The laser power is less than 1 mW to avoid overheating of the samples.

**TEM and STEM Characterization.** The STEM samples were prepared with a poly (methyl methacrylate) (PMMA) assisted method. A layer of PMMA of about 1 μm thick was firstly spin-coated on the wafer with MoTe$_2$ samples deposited, and then baked at 180 °C for 3min. The wafer was then immersed in NaOH solution (1M) overnight to etch the SiO$_2$ layer. After lift-off, the PMMA/MoTe$_2$ film was transferred into distilled (DI) water for several cycles to rinse off the residual contaminants, and then it was fished by a TEM grid (Quantifoil Au grid). The transferred specimen was dried naturally in ambient environment, and then dropped into acetone overnight to dissolve the PMMA coating layers. The STEM imaging on MoTe$_2$ were performed on a JEOL 2100F with a cold field-emission gun and a DELTA aberration corrector operating at 60 kV. A Gatan GIF Quantum was used to record the EELS spectra. The inner and outer collection angles for the STEM images (β1 and β2) were 62 and 129–140 mrad, respectively, with a convergence semi-angle of 35 mrad. The beam current was about 15 pA for the ADF imaging and EELS chemical analyses. All imaging was performed at room temperature.

**Devices fabrication and transport measurement.** Few-layer MoTe$_2$ crystals with the thickness ranging from 2 nm to 30 nm were firstly identified by their color contrast under optical microscopy. Then small markers were fabricated using standard e-beam lithography (EBL) near the identified sample for subsequent fabrication of Hall-bar devices. To obtain a clean interface between the electrodes and the sample, *in situ* argon plasma was employed to remove the resist residues before metal evaporation without breaking the vacuum. The Ti/Au (5/70 nm) electrodes were deposited using an electron-beam evaporator followed by lift-off in acetone. Transport experiments were carried out with a standard four-terminal method from room temperature to 0.3 K in a top-loading Helium-3 refrigerator with a 15 T superconducting magnet. A standard low-frequency lock-in technique was used to measure the resistance with an excitation current of 10 nA. Angular-dependent measurements were facilitated by an *in situ* home-made sample rotator.

**Acknowledgements**
The authors thank Xianxin Wu, Heng Fan, Jianlin Luo, Hsin Lin, and Jiangping Hu for stimulating discussions. This work has been supported by the National Basic Research Program of China from the MOST under the grant No. 2014CB920904, 2015CB921101 and 2016YFA0300600, by the NSFC under the grant Nos. 11527806 and 11874406. Research in Singapore was funded by the Singapore National Research Foundation under NRF award number NRF-RF2013-08 and MOE Tier 2 MOE2016-T2-2-153. J.L. and K.S. acknowledge JST-ACCEL and JSPS KAKENHI (JP16H06333 and P16382) for financial support. J.L. acknowledges financial support from the Hong Kong Research Grants Council (Project No. ECS26302118). K.L. thanks the support of HKRGC and Croucher Foundation.

**Author contributions**
J.C., P.L., J.Z., and W.-Y.H. contributed equally to this work. G.L. and Z.L. conceived and supervised the project, and designed the experiments; J.C., P.L. and X.H. fabricated the devices and carried out the transport measurements; J.Z. synthesized the sample; J.L. did the measurements and data analysis on STEM; W.H. and K.L. predicted the presence of the anisotropic SOC and its experimental implications. J.L. performed the first-principle calculations. G.L. prepared the manuscript with input from Z.L., J.L., J.Z., W.H., K.L., and J.L. All the authors discussed the results and commented on the manuscript.

**Additional information**

Supplementary information is available in the online version of the paper. Reprints and permissions information is available online at www.nature.com/reprints.

Correspondence and requests for materials should be addressed to J.L., Z.L. or G.L.

**Competing financial interests**

The authors declare no competing financial interests.

**Figure and Figure Caption**

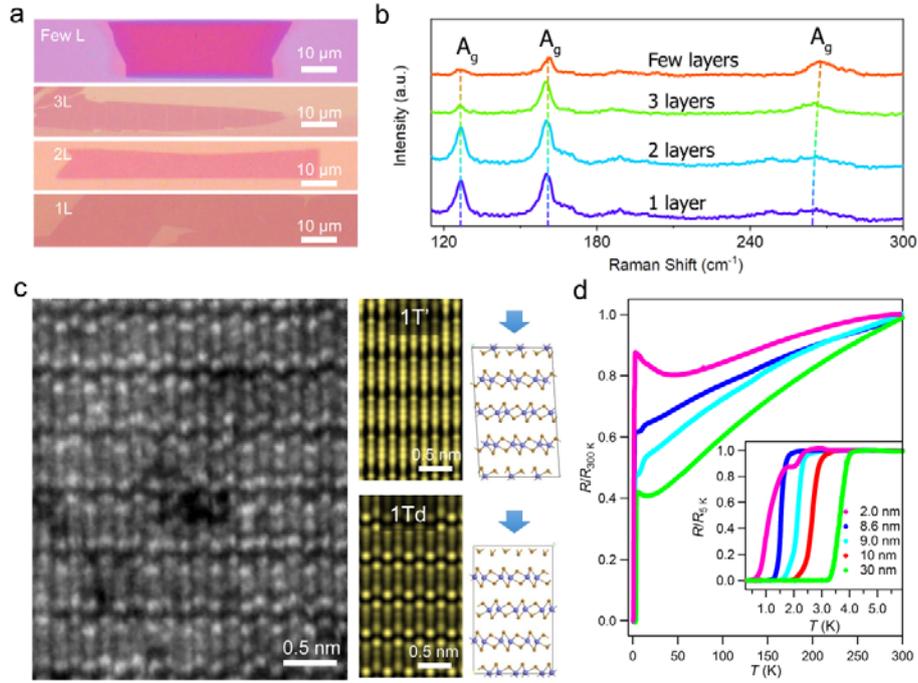

**Figure 1 | Structural and transport characterization of the as-synthesized 1Td-MoTe$_2$ samples**. **a**, Optical images of monolayer (1L), bilayer (2L), trilayer (3L) and few layers of as-synthesized MoTe$_2$. The size of the monolayer sample can reach up to 100 μm. **b**, Raman spectra of the few-layer MoTe$_2$ samples. Raman peaks were observed at 127, 161, and 267 cm$^{-1}$, corresponding to the Ag modes of 1Td-MoTe$_2$. **c**, Atomic resolution STEM image of few-layer 1Td-MoTe$_2$. Simulated STEM images of few-layer MoTe$_2$ in 1T' and 1Td stacking viewed along [001] zone axis are shown next to the experimental image, respectively. Compared with the simulation, the stacking of the few-layer MoTe$_2$ is confirmed to be the 1Td phase. **d**, Superconductivity in few-layer 1Td-MoTe$_2$. The inset shows the temperature dependence of the reduced four-terminal resistance ($R/R_{5K}$) in the range from 0.3 K to 4.5 K, for MoTe$_2$ devices with the thickness ranging from 2 to 10 nm.

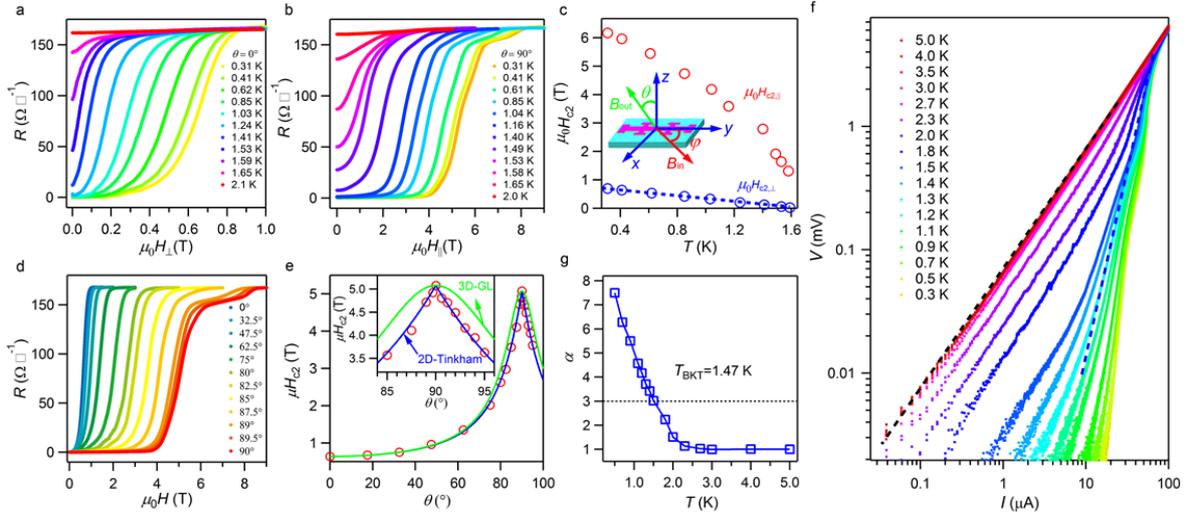

**Figure 2 | Two-dimensional superconductivity in few-layer 1Td-MoTe$_2$ crystals. a**, **b**, Superconducting resistive transition of the 8.6-nm-thick MoTe$_2$ crystal in perpendicular magnetic field (**a**) and in parallel magnetic field (**b**). **c,** Temperature dependence of the upper critical field $\mu_0 H_{c2}$ corresponding to reduced resistance $r = 0.5$, with magnetic field directions parallel ($\mu_0 H_{c2,\parallel}$) and perpendicular ($\mu_0 H_{c2,\perp}$) to the crystal plane. The dashed line is fitting to the 2D Ginzburg-Landau theory. The inset is a schematic drawing of the tilt experiment setup, where $x$, $y$, and $z$ represents the crystallographic $b$, $a$, and $c$ axis, $\theta$ is the out-of-plane tilted angle between the out-of-plane magnetic field $B_{\text{out}}$ and the positive direction of $z$-axis, and $\varphi$ is the in-plane tilted angle between the in-plane magnetic field $B_{\text{in}}$ and the positive direction of $y$-axis. **d,** Magnetic field dependence of the sheet resistance of the 8.6-nm MoTe$_2$ device at $T$=0.3 K with different tilted angles $\theta$. **e,** Angular dependence of the upper critical field $\mu_0 H_{c2}$. The inset shows a zoom-in view of the region around $\theta$=90º. The solid lines represent the fitting with the 2D Tinkham formula $\left|\frac{H_{c2}(\theta)\cos\theta}{H_{c2,\perp}}\right| + \left(\frac{H_{c2}(\theta)\sin\theta}{H_{c2,\parallel}}\right)^2 = 1$ (blue line) and the 3D anisotropic mass model (3D-GL) $\left(\frac{H_{c2}(\theta)\cos\theta}{H_{c2,\perp}}\right)^2 + \left(\frac{H_{c2}(\theta)\sin\theta}{H_{c2,\parallel}}\right)^2 = 1$ (green line), respectively. **f,** Voltage-current behavior (*V-I*) at different temperatures close to $T_c$ plotted on a logarithmic scale. The black

dashed line indicates the ohmic behavior expected in the normal state. The blue dashed line indicates $V \propto I^3$, which corresponds to $T = T_{BKT}$. **g**, Temperature dependence of the exponent $\alpha$ obtained from fittings data in (**e**) to $V \propto I^\alpha$. $T_{BKT}=1.47$ K is obtained from the intersection between the experimental curve and the dashed black line $\alpha=3$.

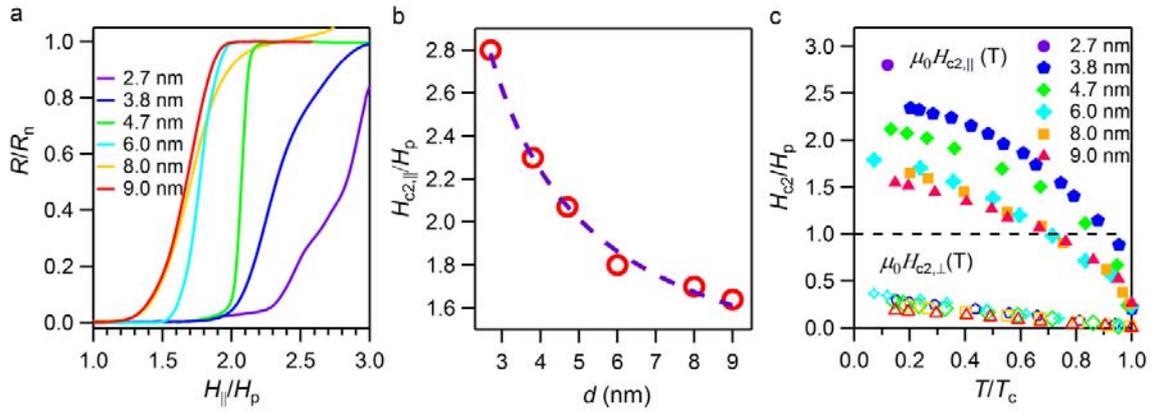

**Figure 3 | Enhanced in-plane upper critical field in few-layer 1Td-MoTe$_2$. a,** Magnetic field dependence of the resistance for 1Td-MoTe$_2$ devices with various thicknesses from 2.7 nm to 9 nm. The resistances and magnetic fields are normalized by the normal state resistance $R_n$ and the Pauli limit $H_p$, respectively. **b**, Normalized in-plane upper critical field $H_{c2,\parallel}/H_p$ as a function of sample thickness $d$. The purple dashed line is a guide to the eye. **c,** Normalized upper critical field $H_{c2}/H_p$ as a function of reduced $T/T_c$ for few-layer MoTe$_2$. The black dashed line denotes the Pauli limit $H_p$.

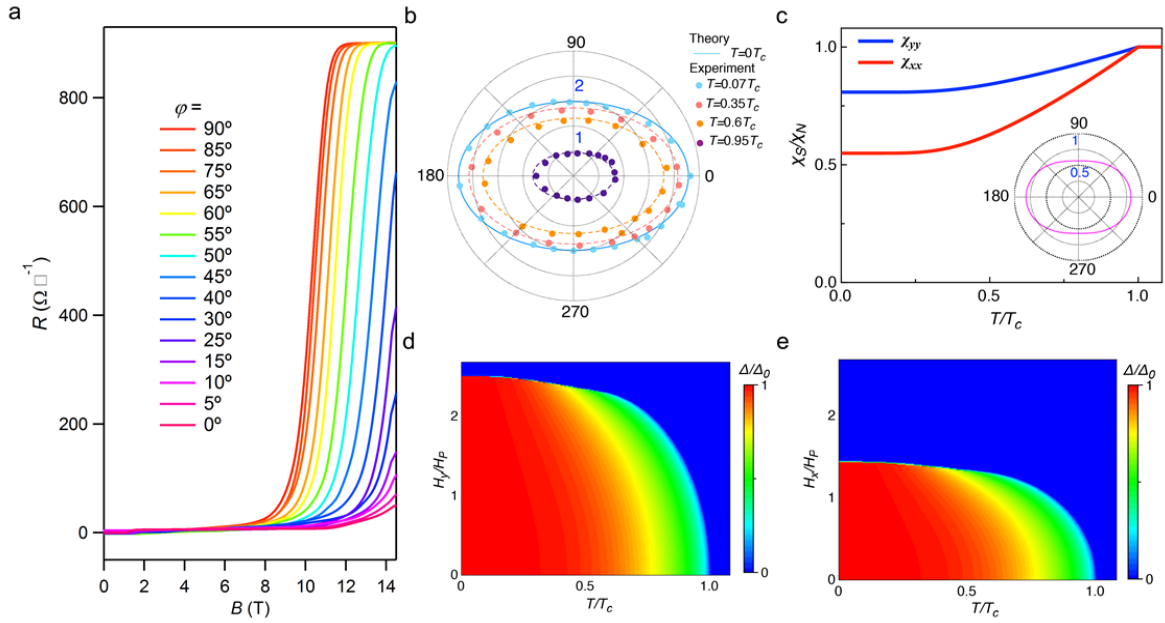

**Figure 4 | Two-fold symmetry of in-plane upper critical field $H_{c2,\parallel}$. a,** Magnetic field dependence of the sheet resistance of the 3-nm-thick $MoTe_2$ device at $T = 0.3$ K ($T = 0.07\ T_c$) with different in-plane tilted angles $\varphi$. **b,** Angular dependence of the in-plane upper critical field normalized by Pauli limit $H_{c2,\parallel}/H_P$. The experimental data are measured at 0.07 $T_c$, 0.35 $T_c$, 0.6 $T_c$, and 0.95 $T_c$. The theoretical value of $H_{c2,\parallel}$ at $T$=0 K is plotted to show the two-fold symmetry consistent with the experimental data at low temperature. The dashed lines are the asymptotic curves to show the two-fold symmetry maintains at $T$=0.35 $T_c$, 0.6 $T_c$, and 0.95 $T_c$. **c,** Temperature dependence of the normalized in-plane spin susceptibility $\chi_S/\chi_N$ along $x$ and $y$ direction respectively. The inset is the polar plot for the zero temperature normalized spin susceptibility. **d,e,** The temperature phase diagram for the superconducting state with anisotropic SOC under $x$ (**d**) and $y$ (**e**) oriented in-plane magnetic field respectively.